\documentclass[aps,prl,twocolumn,floatfix,english,showpacs,10pt,superscriptaddress]{revtex4-2}%
\usepackage{graphicx}
\usepackage{booktabs}
\usepackage{amsmath}
\usepackage{physics}
\usepackage{amssymb}
\usepackage{mathrsfs}
\usepackage{colordvi}
\usepackage{verbatim}
\usepackage{xcolor}
\usepackage{multirow}
\usepackage{mathrsfs}
\usepackage{epsfig}
\usepackage{lipsum}
\usepackage{amsfonts}
\usepackage[unicode=true, breaklinks=false, pdfborder={0 0 1}, backref=false,
colorlinks=true, linkcolor=blue, urlcolor=blue, citecolor=blue]{hyperref}%
\providecommand{\U}[1]{\protect\rule{.1in}{.1in}}

\setcitestyle{numbers,square}
\begin{document}
\title{Non-Hermitian Aharonov-Bohm Cage in Bosonic Bogoliubov-de Gennes Systems}
\author{Kunling Zhou}


\affiliation{School of Physics, Huazhong University of Science and Technology, Wuhan 430074, P. R. China}

\author{Bowen Zeng}
\email[]{zengbowen@csust.edu.cn}
\affiliation{Hunan Provincial Key Laboratory of Flexible Electronic Materials Genome Engineering,
School of Physics and Electronic Sciences, Changsha University of Science and Technology, Changsha 410114, P. R. China}

\author{Yong Hu}
\email[]{huyong@hust.edu.cn}
\affiliation{School of Physics, Huazhong University of Science and Technology, Wuhan 430074, P. R. China}

\makeatletter
\newcommand{\rmnum}[1]{\romannumeral #1}
\newcommand{\Rmnum}[1]{\expandafter\@slowromancap\romannumeral #1@}
\makeatother

\begin{abstract}
The non-Hermitian Aharonov-Bohm (AB) cage is a unique localization phenomenon that confines all possible excitations. This confinement leads to fully flat spectra in momentum space, which are typically accompanied with the degeneracy with various types. Classifying the degeneracy type is crucial for studying the dynamical properties of the non-Hermitian AB cage, but the methods for such classification and their physical connections remain not very clear. Here, we construct a non-Hermitian AB cage in a bosonic Bogoliubov-de Gennes (BdG) system with various types of degenerate flat bands (DFBs). Using the transfer matrix, we demonstrate the localization mechanism for the formation of AB cage and derive the minimal polynomial in mathematics for classifying the degeneracy types of DFBs, thus providing comprehensive understanding of the correspondence among the degeneracy type of DFBs, the minimal polynomial, and the transfer matrix. With such correspondence, we propose a scheme to realize highly degenerate flat bands. 

\end{abstract}

\maketitle

\section{introduction}

Flat bands refer to dispersion relations that are independent of momentum, leading to zero group velocity and the consequent localization of excitations~\cite{PhysRevA.82.041402,PhysRevA.87.023614,PhysRevLett.114.245504,PhysRevLett.114.245503,WOS:000314858600016,PhysRevLett.115.143601,DENG2003412,PhysRevLett.84.143,PhysRevA.96.011802,PhysRevA.99.033810,WOS:001045055600001,PhysRevLett.123.183601,Weimann:16,WOS:000216599300018,PhysRevLett.83.5102,PhysRevA.93.062319}. The Aharonov-Bohm (AB) cage~\cite{PhysRevB.64.155306,PhysRevLett.81.5888,PhysRevLett.121.075502,PhysRevLett.129.220403} is a unique localization phenomenon that confines all possible excitations, thereby giving rise to entirely flat spectra across all bands. This character renders the AB cage a good platform for exploring the strongly correlated physics. The formation of AB cage arises from the complete destructive interference induced by the interplay between an external gauge field and the lattice geometry~\cite{Longhi:14,PhysRevLett.85.3906,PhysRevLett.81.5888}. A recent study~\cite{PhysRevA.102.023524} introduced a novel non-Abelian AB cage, in which the condition for the destructive interference is generalized to the nilpotent interference matrix. This generalization enables different AB cages to be characterized by different nilpotent indices that are governed  by the order of exceptional points (EPs), a unique spectral degeneracy phenomenon with the collapse of the eigenstates space~\cite{doi:10.1080/00018732.2021.1876991,yu2024non} in the non-Hermitian systems.
Thus, the formation of an AB cage  generally leads to the flat bands with degeneracy, which are referred to as degenerate flat bands (DFBs) in this work.

The degeneracy in DFBs endows the AB cage with abundant physics such as enhanced sensitivity to perturbation arising from the nonlinear spectral structure in the vicinity of the EPs~\cite{chen2017exceptional,hodaei2017enhanced,hokmabadi2019non,doi:10.1021/acsphotonics.1c01535}. Recently, the energy degeneracy associated with high-order EPs has attracted much attention ~\cite{PhysRevResearch.4.023130,PhysRevLett.127.186601,doi:10.1126/sciadv.adi0732,PhysRevLett.127.186602} and there is a growing demand for DFBs with higher degeneracy degrees~\cite{PhysRevB.111.115105,PhysRevB.106.165133,becker2025degenerateflatbandstwisted,PhysRevLett.126.106601,PhysRevLett.132.246401}. One of the driving factors behind this is that additional filling selections of highly degenerate flat bands pave the way for a broader range of strongly correlated physics phenomena~\cite{PhysRevLett.132.246401}. DFBs with higher degeneracy degrees usually exhibit various types of degeneracy, which can manifest as the diabolical points (DPs) type, the EPs type or a combination of them~\cite{PhysRevA.108.023518,9146199,PhysRevResearch.2.033127,Zhang_2024,10.1063/5.0174456}. Classifying these degeneracy types holds great significance.
For example, for DFBs with different types, the excitations transition in different ways, causing the AB cage to exhibit distinct dynamical properties and local structures~\cite{WOS:000428961400003,PhysRevA.102.023524}. A mathematical approach to classify the degeneracy type of DFBs involves utilizing the minimal polynomial of the matrix~\cite{garcia2017second}, but which lacks physical intuition. It is hypothesized that the potential physical connections might be embedded within the transfer matrix~\cite{bernevig2013topological,Mostafazadeh2020TransfermatrixIS}, which represents the transition probability between two states. However, such connections have yet to be fully understood.

To comprehensively illustrate the relationship among the degeneracy type of DFBs in the AB cage, the minimal polynomial, and the transfer matrix, we choose the bosonic Bogoliubov–de Gennes (BdG) system as a specific example which has been realized experimentally in various platforms ranging from superconduct circuits~\cite{busnaina2024quantum}, magnonic systems~\cite{nawa2019triplon,10.1093/ptep/ptaa151}, optomechanical systems~\cite{del2022non,slim2024optomechanical} to hybird systems~\cite{marti2024quantum}. Experimentally, the gain-loss systems via coupling to bath are more commonly used to investigate the properties of non-Hermitian degeneracies~\cite{miri2019exceptional}. However, the presence of bath-induced noise and the necessity for post-selection protocols pose challenges in observing pure non-Hermitian effects~\cite{PhysRevLett.127.140504,PhysRevA.108.032214,lau2018fundamental}. In contrast, the inherent non-Hermitian nature of BdG dynamical matrices facilitates the manifestation of non-Hermitian dynamics without involving gain or loss~\cite{PhysRevA.99.063834,10.1093/ptep/ptaa151,PhysRevB.103.165123,PhysRevX.8.041031}, which enables the BdG system to serve as a good platform for observing the non-Hermitian AB cage~\cite{busnaina2024quantum,kang2025dissipation}. Moreover, the inherent particle-hole symmetry and pseudo-symmetry~\cite{PhysRevLett.130.203605,10.1063/5.0035358} 
in BdG systems enable the realization of DFBs with high degeneracy degrees.

In such a non-Hermitian AB cage constructed in bosonic BdG system, by tuning the system parameters, we demonstrate a flexible control over the degeneracy type of flat bands, ranging from DPs type to higher-order EPs type. Unlike the conventional method of distinguishing degeneracy types based on the response to external perturbations, here we establish the correspondence between parameter-dependent transfer matrix and the minimal polynomial that can be used to determine the degeneracy types of DFBs. Meanwhile, we also uncover the localized mechanism for the formation of AB cage by investigating the limitation of the transfer matrix on the prohibited propagation paths. Applying such correspondence to the system with $N$ coupled chains, the transfer matrix can be designed to achieve DFBs with $2N$-order EPs type degeneracy.

\section{model}
\begin{figure}
    \centering
    \includegraphics[width = 8.6 cm]{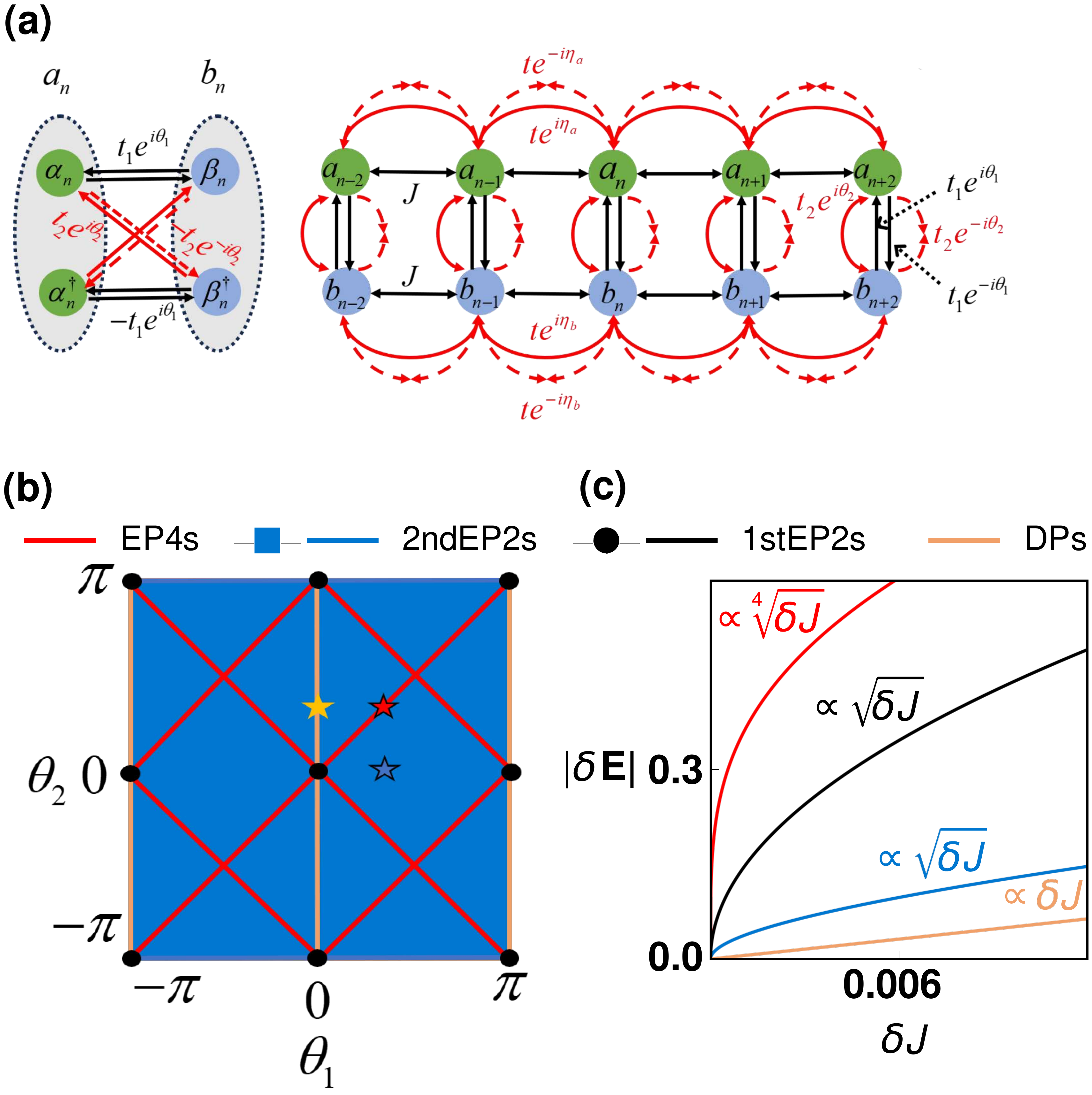}
    \caption{(a) Schematic diagram of the ladder model, consisting of coupled two Kitaev-Majorana chains (right side), where the conjugated coupling and two-boson creation/annihilation processes are denoted by the black lines, red solid/dashed lines, respectively. Corresponding to these couplings, the hopping strengths of the particle and hole degrees of freedom between $a_n$ and $b_n$ is shown in the left side. Under a transformation that exchanges the particle and hole [i.e. $\alpha(\beta)\rightarrow \alpha^{\dagger}(\beta^{\dagger})$,  $\alpha^{\dagger}(\beta^{\dagger})\rightarrow \alpha(\beta)$], the coupling strength acquires an anti-conjugated form  thereby manifesting the particle hole symmetry in BdG systems~\cite{10.1063/5.0035358}.  (b) The $ \theta_1$-$\theta_2$ phase diagram of degeneracy type of DFBs, where the red lines, yellow lines, blue regions and black dots correspond to EP4s type, DPs type, 1stEP2s type and 2ndEP2s type in the energy band. 
    (c) The dependence of the absolute value of eigenvalues variation on the perturbation $\delta J $ for different types of DFBs, when $J = t = 2$ and the wave vector is chosen as $k=0$.  The parameter choices are $\left\{ t_1=t_2=2, \theta_1=-\theta_2=\pi/3 \right\}$ for the EP4s type,   $\left\{t_1=2\sqrt{3}, t_2=2, \theta_1=\pi/3,\theta_2=\pi/6 \right\}$  for the 2ndEP2s type,  $\left\{t_1=t_2=2, \theta_1=\theta_2=0\right\}$ for the 1stEP2s type, and $\left\{t_1=1, t_2=2, \theta_1=0, \theta_2=\pi/3\right\}$ for the DPs type. 
    The $\theta_1,\theta_2$ for the 1stEP2s type are located at the origin point, while those for the other types are marked with pentagrams in their corresponding colors in (b).  }
    \label{fig-model}
\end{figure}

We consider a ladder model consisting of two identical bosonic Kitaev-Majorana chains, denoted as chain $``a"$ and chain $``b"$, as shown in Fig.~\ref{fig-model}(a). The conjugated coupling (black lines) and number non-conservation coupling including two-boson creation and annihilation processes (red solid and dashed lines) are parameterized by $J$ and $te^{\pm i \eta_{\left(a, b\right)}}$  for chains $(``a", ``b")$. Their counterparts for the rungs of the ladder are parameterized by  $ t_1 e^{\pm i\theta_1} $ and $ t_2 e^{\pm i\theta_2} $, respectively. The  Hamiltonian for this system is given by
\begin{align}
    H=&\sum_n J  \alpha^{\dagger}_n \alpha_{n+1} +J  \beta^{\dagger}_n \beta_{n+1}+t_1 e^{i\theta_1} \alpha^{\dagger}_n \beta_n \nonumber \\
    +&t_2 e^{i\theta_2} \alpha_n^{\dagger} \beta_n^{\dagger} +t e^{i\eta_a} \alpha^{\dagger}_n \alpha^{\dagger}_{n+1}+t e^{i\eta_b} \beta^{\dagger}_n \beta^{\dagger}_{n+1}+\text{H.c.},    
\end{align}
where $\left(\left\{\alpha_n,\alpha_n^{\dagger} \right\}, \left\{\beta_n,\beta_n^{\dagger} \right\}   \right)$ denote the creation and annihilation operators for chains $(``a", ``b")$ on $n$-site. Without loss of generality, we choose $t_1$,  $t_2 $, $t$, and $J$ as real numbers, while $ \eta_a$, $\eta_b$, $\theta_1$, and  $\theta_2$  are phase factors related to the gauge choice. Considering the gauge invariance for non-Abelian gauge field~\cite{asboth2016short,makeenko2009brief}, two gauge degrees of freedom remain, as detailed in the Appendix~\ref{Appendix-A}. Consequently, we can always choose a gauge where $\eta_a = 0$  and  $\eta_b = 0$, while keeping phase factors $\theta_1$ and $\theta_2$. In this gauge, the Hamiltonian in momentum space is expressed as
\begin{align}
\label{inter leg phase gauge H}
   &H(k)=\sum_k\psi(k)^{\dagger}M(k)\psi(k)  \nonumber \\
  = &\sum_k 2J\cos{k} (\alpha_k^{\dagger} \alpha_k + \beta_k^{\dagger} \beta_k)
   +(t_1 e^{i\theta_1} \alpha^{\dagger}_k \beta_k +t_2 e^{i\theta_2} \alpha^{\dagger}_k \beta^{\dagger}_{-k} \nonumber \\
   &+t \cos{k} \alpha_k^{\dagger}\alpha_{-k}^{\dagger} +t  \cos{k} \beta_k^{\dagger}\beta_{-k}^{\dagger} +\text{H.c.}) .
\end{align}
Here, $\psi(k)=\qty[\alpha_k,\beta_{k},\alpha_{-k}^{\dagger},\beta_{-k}^{\dagger}]^\intercal$ and $M(k)$ is the matrix form of $H(k)$ under the basis $\psi(k)$. In the bosonic BdG systems, the evolution of mode $\psi(k)$ is governed by the Heisenberg equation of motion, 
\begin{align}
    i \frac{\mathrm{d} \psi(k)}{\mathrm{d} t}=[\psi(k), H(k)]=\tilde{H}(k)\psi(k), 
\end{align}
where $\tilde{H}(k)$ is usually referred to as the dynamical matrix. In terms of Bogoliubov transformation, the dynamical matrix is associated with $M(k)$ in Eq.~\eqref{inter leg phase gauge H} by~\cite{PhysRevB.89.054420,PhysRevB.98.115135,PhysRevB.103.165123,xiao2009theorytransformationdiagonalizationquadratic,COLPA1978327}
\begin{align}
    \tilde{H}(k)=\gamma_0 M(k).
\end{align} 
Here, $\gamma_0=\sigma_3 \otimes I_2$ with $\sigma_3$ being the Pauli matrix and $I_n$ being the $n \times n$ identity matrix. A conventional approach for solving the eigenvalues of $\tilde{H}(k)$ is to express $\tilde{H}(k)$ in terms of Dirac gamma matrix in Dirac representation~\cite{marsh2017mathematics}, 
\begin{align}
\label{eq-Hamiltonian-gammamatrix}
    \tilde{H}(k)&=2J\cos k \gamma^0+t_1\cos \theta_1 \gamma^1 \gamma^5+\mathrm{i}t_1\sin \theta_1 \gamma^1\gamma^3 \notag \\
    &+2t\cos k\gamma^0\gamma^5+t_2\cos\theta_2\gamma^1+\mathrm{i}t_2\sin \theta_2 \gamma^0\gamma^1.
\end{align}
Here $\gamma^i = i\sigma_2 \otimes \sigma_i$ for $i=1,2,3$, and $\gamma^5=\sigma_1 \otimes I_2$ with $\sigma_i$ being the Pauli matrix. Utilizing the properties of $\gamma$ matrix and squaring two sides of the equation twice yield the annihilating polynomial~\cite{garcia2017second} of $\tilde{H}(k)$ as
\begin{align}
\label{periodic dynamical eigenvalues}
    (\tilde{H}(k)^2-(\lambda-\sqrt{\delta})I_{4})(\tilde{H}(k)^2-(\lambda+\sqrt{\delta})I_{4})=0
\end{align}
with parameters $\lambda^2 = t_1^2 - t_2^2 + 4(J^2 - t^2)\cos^2 k $ and $\delta = 4\cos^2 k \left( (Jt_1 \cos \theta_1 - tt_2 \cos \theta_2)^2 + (J^2 - t^2) t_1^2 \sin^2 \theta_1 \right)$. According to the
Hamilton-Cayley theorem ~\cite{garcia2017second,hungerford2012algebra,jacobson2012basic}, the annihilating polynomial Eq.~\eqref{periodic dynamical eigenvalues} is identical to the character polynomial and the roots $\pm \sqrt{\lambda \pm 2\sqrt{\delta}}$ serve as the eigenvalues of $\tilde{H}(k)$. 
It is clear that when $t=J$ and $t_1 \cos \theta_1 = t_2 \cos \theta_2$, the dispersion relation of $\tilde{H}(k)$ is simplified to $\pm\sqrt{t_1^2 - t_2^2}$ independent of momentum, exhibiting a non-Hermitian AB cage with completely flat bands. 

Such flat bands are degenerated involving at least two degeneracy degrees. Under the constraint $ t_1 \cos \theta_1 = t_2 \cos \theta_2$, Fig.~\ref{fig-model}(b) shows the $\theta_1$-$\theta_2$ phase diagram for DFBs. Our system allows four types of degeneracy, including a DPs type, two kinds of EP2s types, and a fourth-order EPs (EP4s) type, as shown in Fig.~\ref{fig-model}(b). 
While previous studies focused on DFBs associated with single type of degeneracy such as EP2s~\cite{10.1063/5.0174456,Zhang_2024,PhysRevResearch.2.033127}, 
 EP3s~\cite{9146199}, and EP4s~\cite{Zhang_2024}, our work unifies these types within a single system and reveals the correspondence between the degeneracy types and the transition properties of the AB cage, as discussed in Section.~\ref{sec4}.  The system exhibits DPs type DFBs when $\theta_1 = n_1 \pi$,  $\theta_2 \neq n_2 \pi$ and $n_1,n_2 \in \mathbb{Z}$ with $E=\pm\sqrt{t_1^2-t_2^2}$, as verified by the linear dependence of the absolute value of eigenvalues on perturbation $\delta J$ [see Fig.~\ref{fig-model}(c)]. The realization of EP4s requires $ \theta_1 \neq  n_1\pi$  and $\theta_1 \pm \theta_2 = n_2 \pi$  with $E=0$. In this case, under perturbation $\delta J$, the eigenvalues variation $\delta \abs{E} \propto \sqrt[4]{\delta J}$, as shown in Fig.~\ref{fig-model}(c).  Notably, a special case $\theta_1=(2n_1+1)\pi/2$ and $\theta_2=(2n_2+1)\pi/2$ is excluded from the EP4s classification in Fig.~\ref{fig-model}(b). Since when $\cos{\theta_1}=\cos{\theta_2}=0$, no constraints are imposed on $t_1$ and $t_2$. Although the phase relationship $\theta_1 \pm \theta_2 = n_2 \pi$ remains satisfied, the system exhibits EP4s only when $\abs{t_1}=\abs{t_2}$ with $E=0$, and any other values of $t_1$ and $t_2$ yield 2ndEP2s. There are two kinds of EP2s in Fig.~\ref{fig-model}(b), where the first EP2s (1stEP2s) type emerges at the intersection of EP4s lines and DPs lines and the second EP2s (2ndEP2s) lies within blue regions. The absolute value of eigenvalues for both two EP2s types exhibit $\sqrt{\delta J}$ with different constant factors. For each type of DFBs, the minimum polynomial and excitation transition varies, as will be discussed later. 

\section{The localization mechanism of the non-Abelian AB cage}

\begin{figure}[h]
    \centering
\includegraphics[width=0.5\textwidth]{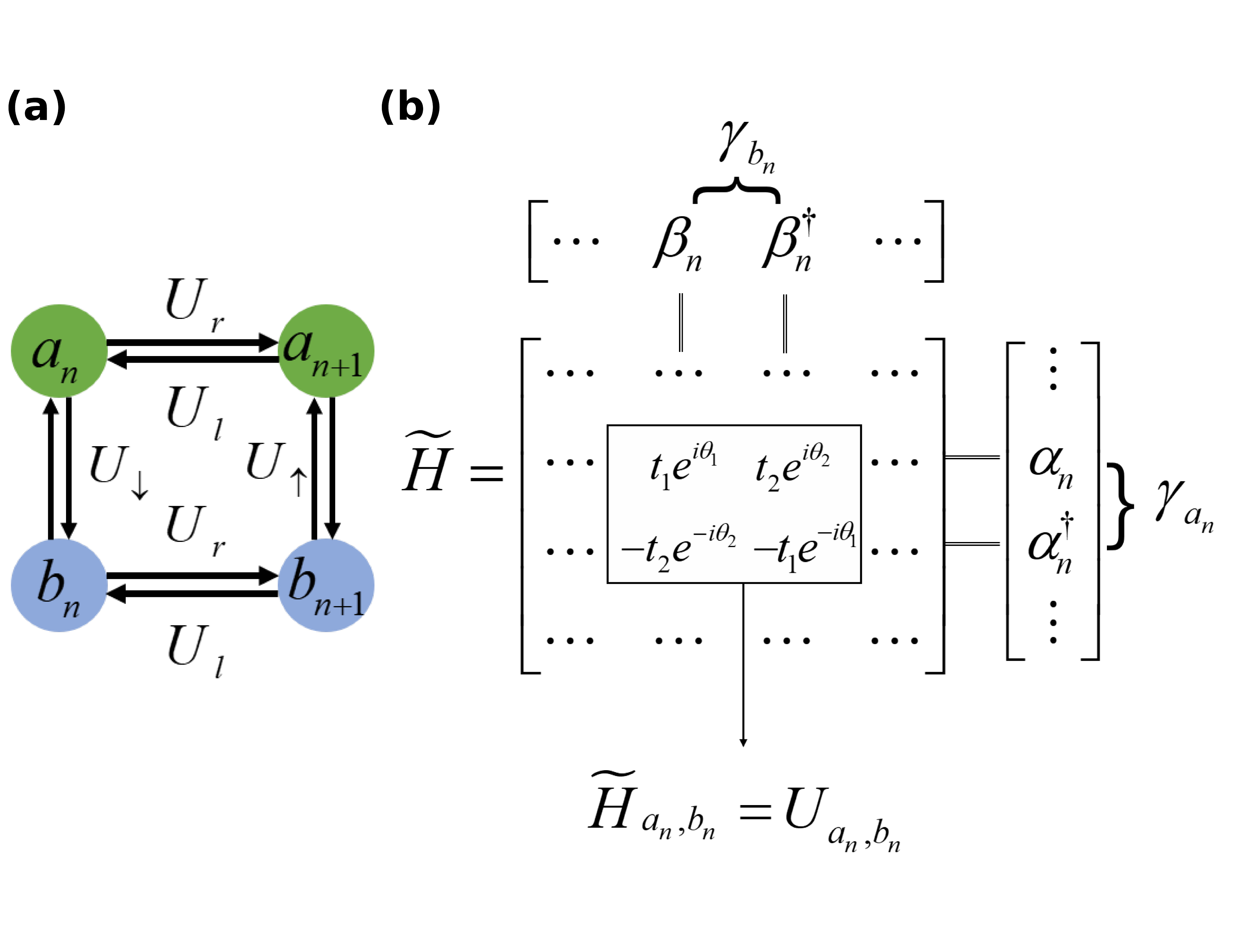}
    \caption{(a) The transfer matrix $U$ between different sites, where $U_l$ ($U_r$, $U_{\uparrow}$, $U_{\downarrow}$) denotes the leftward (rightward, upward, downward) transition. (b) The dynamical matrix $\tilde{H}$ in real space involving all possibilities of propagation, which is exemplified by a specific transfer matrix $U_{a_n,b_n}$ from $b_n$ to $a_n$. }
    \label{fig2}
\end{figure}

To provide insights behind rich degeneracy types of DFBs in our system and uncover the localization mechanism of the non-Abelian AB cage in real space, we consider an excitation in the ladder model and trace its transition using the transfer matrix, as shown in Fig.~\ref{fig2}. 

We firstly elucidate the method of transfer matrix used here. In our system, the particle and hole degrees of freedom on a single site can be expressed as a two-component operator $ \gamma_{a_n(b_n)} = [  \alpha_n(\beta_n), \alpha^{\dagger}_n(\beta^{\dagger}_n)   ]^{\intercal}$. The Heisenberg equation of motion for the operator $\gamma_{a_n}$ is thus given by
\begin{align}
\label{eq-non-abelian}
    i\dot{\gamma}_{a_n} =  \sum_m \left( U_{a_n,a_m} \gamma_{a_m}+ U_{a_n,b_m} \gamma_{b_m} \right).
\end{align}
Here, both of $U_{a_n,a_m}$ and  $U_{a_n,b_m}$ are the $2\times2 $ transfer matrices that characterize the transition probability from site $ a_m$ and $b_m$ to site $a_n$. A similar analysis applies to $\gamma_{b_n}$. Now, the conjugate coupling and number non-conserving coupling between two sites in Fig.~\ref{fig-model}(a) can be replaced by these transfer matrices, which are similar to the  translational-invariant link variables produced by a non-Abelian gauge field~\cite{PhysRevA.102.023524,makeenko2009brief}. From this perspective, these transfer matrices typically do not commute with each other. Due to the translation symmetry of system, we can use four types of transfer matrices $U_r$, $U_l$, $U_{\uparrow}$ and $U_{\downarrow}$ to represent the transition between the neighbor sites where the subscript \{$r$,$l$,$\rightarrow$, $\uparrow$\} denote the transition directions, as depicted in Fig.~\ref{fig2}(a) and these transitions are thus non-Abelian. Here, $U_r$ and $U_l$ have the form of
\begin{align}
    U_r = U_{a(b)_{n+1},a(b)_{n}} = \mqty[&J  &t \\&-t &-J] \nonumber \\
    U_l = U_{a(b)_{n-1},a(b)_{n}} = \mqty[&J &t \\&-t &-J ],
\end{align}
representing the transfer matrices for the rightward and leftward propagation along the chain $``a"$ and $``b"$. $U_{\uparrow}$ and $U_{\downarrow}$ are given by
\begin{align}
    &U_{\uparrow} = U_{a_n,b_n}= \mqty[&t_1\mathrm{e}^{\mathrm{i}\theta_1} &t_2\mathrm{e}^{\mathrm{i}\theta_2} \\ &-t_2\mathrm{e}^{-\mathrm{i}\theta_2}  &-t_1\mathrm{e}^{-\mathrm{i}\theta_1}] \nonumber\\  
    &U_{\downarrow} = U_{b_n,a_n}= \mqty[&t_1\mathrm{e}^{-\mathrm{i}\theta_1} &t_2\mathrm{e}^{\mathrm{i}\theta_2} \\ &-t_2\mathrm{e}^{-\mathrm{i}\theta_2}  &-t_1\mathrm{e}^{\mathrm{i}\theta_1}],
\end{align} 
representing the transfer matrices for the upward and downward propagation between the chain $``a"$ and $``b"$.

The transfer matrix $ U_{a(b)_m,a(b)_n} $ corresponds to the sub-matrix $\tilde{H}_{a(b)_m,a(b)_n} $ of dynamical matrix $\tilde{H}$ under the basis $\ket{\psi} = [\cdots, \gamma_{a_n}, \gamma_{b_n}, \cdots]^{\intercal}$, as shown in Fig.~\ref{fig2}(b). For the transition probability from site $z_l$ to site $z_0$ after $l$ hops, we can further introduce a notation $U^l_{z_0,z_l} = \sum_{z_1,z_2, \cdots,z_{l-1}}  U_{z_0,z_1}U_{z_1,z_2} \cdots  U_{z_{l-1},z_l}$, where $z_1,z_2, \cdots, z_{l-1}$ are the sites experienced during these transitions, and the summation involves all possible paths. Similarly, $U^l_{x_m,y_n}$ is the sub-matrix $\tilde{H}^l_{x_m,y_n}$ of $\tilde{H}^l$, where $\tilde{H}^l$ is $\tilde{H}$ to the power of $l$.

The transfer matrix plays a key role in determining the confined area of the AB cage and the associated flat bands as shown below. The flat bands are usually formed by destructive interference~\cite{PhysRevLett.114.245504,PhysRevLett.114.245503}. Here, the destructive interference of different transition paths can be effectively captured by the transfer matrix as $U^l_{z_0,z_l} = 0$. Additionally, a single propagation path $U_{z_0,z_1}U_{z_1,z_2} \cdots  U_{z_{l-1},z_l} = 0$ can also confine the excitation through interference by the internal degrees of freedom within the transfer matrix.

\begin{figure}[h]
    \centering
   \includegraphics[width=0.46\textwidth]{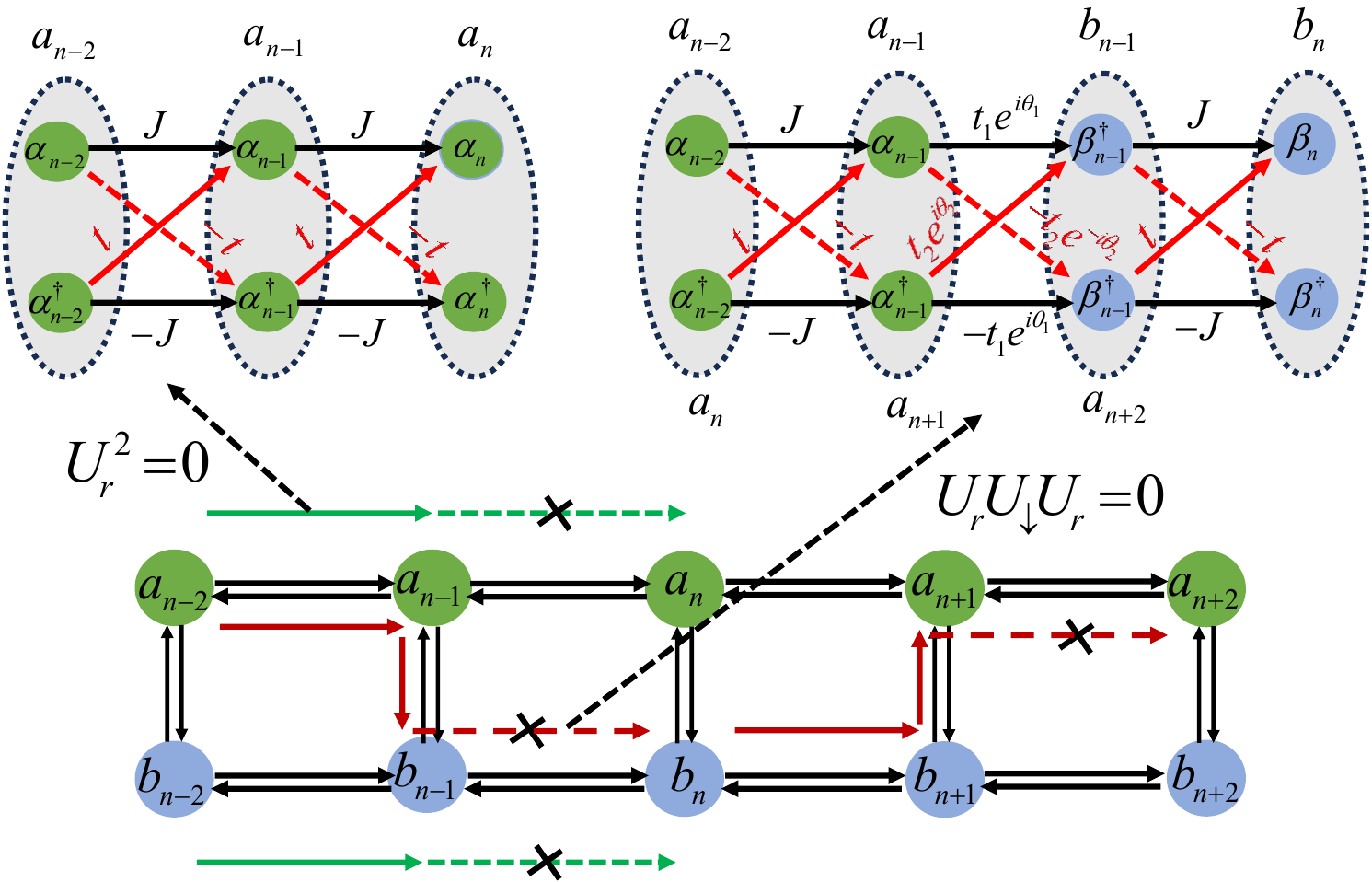}
    \caption{Schematic diagram of the localization mechanism for the formation of DFBs, which can be simplified as the excitation at the $n$-th column cannot hop to the $(n\pm 2)$-th column along the red and green propagation paths. In each path, the destructive interference occurs within the transfer matrix.}
    \label{fig3}
\end{figure}

The sufficient conditions for our system to form an AB cage is that the excitation at the $n$-th column cannot reach the $\left(n \pm 2\right)$-th column. These conditions can be simplified as two limited single propagation paths, as marked by the green and red lines in Fig.~\ref{fig3}. 

To prevent transmission along the green propagation path, we require $U_l^2 = U_r^2 = 0$, which yields $J = t$ and $U_l=U_r$. For simplicity, we introduce the notation $U_0$ to represent both $U_r$ and $U_l$. This prevention can also be regarded as the destructive interference by internal degrees of freedom within the transfer matrix, as shown in the upper part of Fig.~\ref{fig3}. For example, the transmission probability from $\alpha_{n-2}$ to $\alpha_{n}$ can be expressed as a superposition of two components, which comprises $-t^2$ corresponding to the path that traverses $\alpha_{n-1}^{\dagger}$ and $J^2$ associated with the path passing through $\alpha_{n-1}$. Correspondingly, the transmission probability of destructive interference through different paths from $\{\alpha_{n-2},\alpha_{n-2}^{\dagger}\}$ to $\{\alpha_{n},\alpha_{n}^{\dagger}\}$ is captured by $U_0^2=0$. 

The prohibition of the red paths requires the condition that
\begin{align}
\label{eq-transion-path-between-chains}
&U_0U_{\uparrow}U_0 = U_0U_{\downarrow}U_0 \nonumber \\
= &2j(t_1\cos\theta_1 - t_2\cos\theta_2)U_0=0,
\end{align}
i.e., $t_1 \cos\theta_1 = t_2 \cos\theta_2$. Similarly, $U_0U_{\downarrow}U_0=0$ also represents destructive interference through many paths from $\{\alpha_{n-2},\alpha_{n-2}^{\dagger}\}$ to $\{\beta_{n},\beta_{n}^{\dagger}\}$ as shown in the upper right corner of Fig.~\ref{fig3}.

It can be seen that the conditions for these two prohibited paths are precisely those for the flat band discussed in the previous section. This is because the flat band in momentum space arises from the constrained excitations in real space. In the Appendix~\ref{Appendix-B}, we also demonstrate that these sufficient conditions are, in fact, necessary as well.

\section{Degeneracy type of the DFB}
\label{sec4}

In this section, we clarify the degeneracy type of the DFBs using the transfer matrix. The underlying logic is that the degeneracy type of the eigenvalues is generally determined by the multiplicity of roots in the minimal polynomial, while the transfer matrix, as the sub-matrix of $\tilde{H}$, can be used to derive the minimal polynomial of $\tilde{H}$.

Now we consider an arbitrary excitation $\ket{\psi}_e$ and investigate the transitions with transfer matrix. Due to the inherent local nature within the AB cage, the excitation transition either terminates at the end point of transition path or it returns to a linear combination of previously visited states after the $m$ times transition. Both of these two cases can be expressed as
\begin{align}
\label{eq-repeatpath}
    \tilde{H}^m \ket{\psi}_e + \sum_{n=0}^{m-1} f_n \tilde{H}^n \ket{\psi}_e=0.
\end{align}
The coefficients $f_n$ in the first case are zeros. The left side of the above equation is defined as the minimal polynomial of $\ket{\psi}_e$ in literature~\cite{jacobson2012basic,curtis2012linear}, which annihilates the state with minimal degree $m$.  Equation~\eqref{eq-repeatpath} shows that for an arbitrary excitation $\ket{\psi}_e$, the experienced linearly independent states $\{\ket{\psi}_e, \cdots, \tilde
{H}^{m-1}\ket{\psi}_e\}$ form a cyclic subspace with dimension $m$~\cite{hungerford2012algebra}, implying that the evolution of $\ket{\psi}_e$ remains within this space. Thus, we can define this dimension $m$ as the local range of the excitation $\ket{\psi}_e$. For excitations at different sites, the local ranges may take different values, and we denote the largest local range of excitations as the local range of AB cage. 

Different from the Eq.~\eqref{eq-repeatpath}, the minimal polynomial of $\tilde{H}$ annihilates the $\tilde{H}$ with minimal degree $l$~\cite{garcia2017second}, which is defined purely in terms of the matrix and is independent of the state, 
\begin{align}
\label{eq-minipoly}
     \tilde{H}^l+\sum_{n=0}^{l-1} c_n \tilde{H}^n = 0, 
\end{align} 
with undetermined coefficients $c_n$. From another perspective, Eq.~\eqref{eq-minipoly} is satisfied for arbitrary excitation. Consequently, the order $l$ should be the upper bound of the local range. In fact, the local range of the AB cage can always take its upper bound, i.e., the order of minimal polynomial, as proved in the Appendix~\ref{Appendix-C}. 

Although under different parameters, the minimal polynomial may take different forms, we can find an annihilating polynomial (e.g. the characteristic polynomial of $\tilde{H}$), which is the multiple of the minimal polynomial~\cite{garcia2017second} across the complete parameter space of non-Hermitian AB cage. Given that each unit cell contains four degrees of freedom, the system can have at most four distinct flat bands. The order of the minimal polynomial of $\tilde{H}$ or the local range of AB cage thus cannot exceed $4$  (the number of eigenvalues), and such annihilating polynomial must have a degree no less than $4$.  Using the transfer matrix, we can obtain different orders $\tilde{H}^n$, as detailed in the Appendix~\ref{Appendix-D}. A combination of $\tilde{H}^n$ up to the fourth degree yields the annihilating polynomial
\begin{align}
\label{eq-general annihilating polynomial}
    (\tilde{H}-\lambda I)^2(\tilde{H}+\lambda I)^2=0,
\end{align}
with $\lambda=\sqrt{t_1^2-t_2^2}$. 

When $l=4$ in Eq.~\eqref{eq-minipoly}, the minimal polynomial is equivalent to the annihilating polynomial while in the other cases the minimal polynomial with $l\leq 4$ can be the factor of the annihilating polynomial~\cite{garcia2017second}. Under specified parameters, the propagation of excitation in systems can destructively interfere at different sites which leads to different local range of AB cage and minimal polynomial mathematically.

\begin{table*}[ht]
     \caption{The intrinsic connection between the degeneracy type of DFBs, the minimum polynomial of $\tilde{H}$ (first column), and the local range of AB cage (second column). The third column shows schematic diagram of the local range with the transition of initial excitation on $a_n$ (orange color), where the arrows of different colors with numerical labels $n$ indicate the $n$-th transition with the corresponding parameter conditions are shown above. The conditions  $t_1\cos\theta_1 = t_2 \cos\theta_2$, $j=t$ must always be held to form the AB cage. The local range is further verified by the numerical results of dynamical evolution under an initial state $ \gamma_{a_n} = [  1/\sqrt{2},1/\sqrt{2}  ]^{\intercal}$, as shown in the fourth column, where the color bar represents the intensity $\abs{\gamma_{a_n(b_n)}}^2$ of wave function. The parameters for the numerical calculations are the same as those in Fig.~\ref{fig-model}(c).}
   \label{table1}
  \begin{tabular}{cccc}
   \toprule[1pt]
  \makebox[0.25\textwidth][c]{ Minimum polynomial   }\hfill \makebox[0.12\textwidth][c]{Local range  }&\makebox[0.3\textwidth][c]{ Excitation transition }&\makebox[0.35\textwidth][c]{Dynamical evolution}\\
    \midrule[0.5pt]   \begin{minipage}[m]{.25\textwidth} \vspace{40pt}  DP2s:  \\ \vspace*{20pt}
    $(\tilde{H}-\lambda)(\tilde{H}+\lambda)$ 
    \end{minipage} %
   \hfill
   \begin{minipage}[m]{.12\textwidth} 
   \vspace{35pt}
   2
   \end{minipage}& \begin{minipage}[m]{.3\textwidth} \vspace{10pt}
    $\theta_1= n_1 \pi$, $\theta_2 \neq n_2\pi$\\
    \vspace{5pt} $\abs{t_1} \neq \abs{t_2}$
    \vfill 
    \includegraphics[width=4cm]{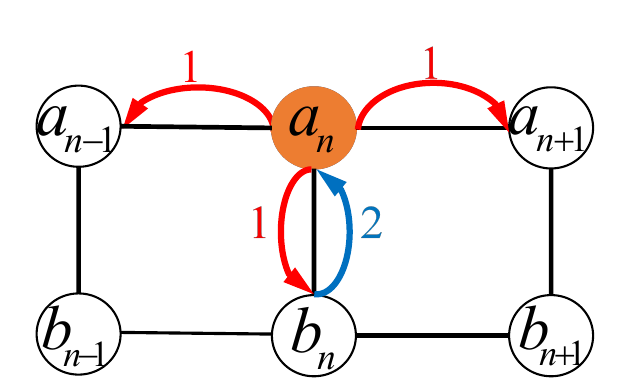}\vspace*{-32pt}
    \end{minipage} & \begin{minipage}[t]{.35\textwidth}
    \centering\vspace*{-25pt}
    \includegraphics[width=5.5cm]{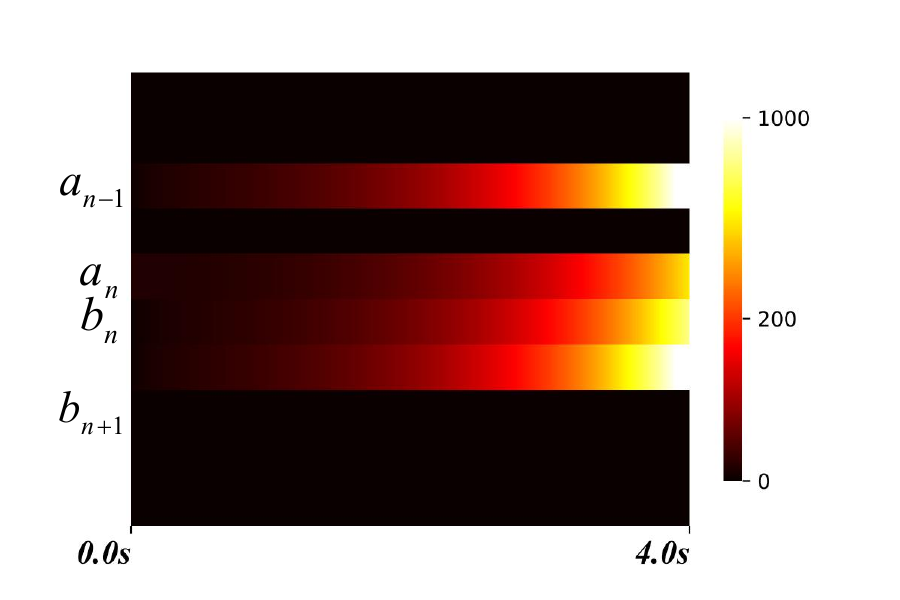}\vspace*{0pt}
    \end{minipage} \\
    
    \midrule[0.5pt]
    \begin{minipage}[m]{.25\textwidth}  
    \vspace{40pt}  1stEP2s:  \\ \vspace*{20pt}
    $\tilde{H}^2$ 
    \end{minipage} %
    \hfill
    \begin{minipage}[m]{.12\textwidth} \vspace{35pt}
    2
    \end{minipage} & \begin{minipage}[m]{.3\textwidth} \vspace{10pt}
    $\theta_1= n_1 \pi$, $ \theta_2 =n_2 \pi$\\
    \vspace{5pt}  $\abs{t_1}=\abs{t_2}$
    \vfill 
    \includegraphics[width=4cm]{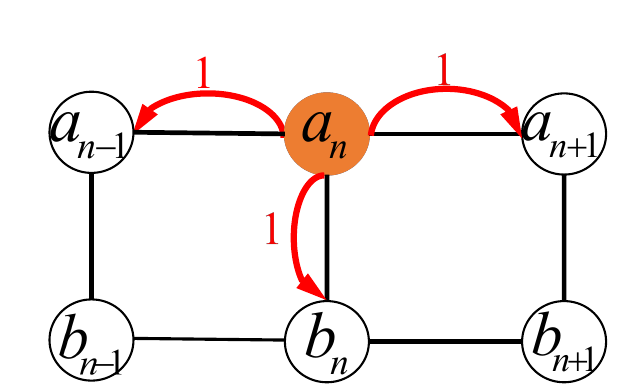}\vspace*{-32pt}
    \end{minipage} & \begin{minipage}[t]{.35\textwidth}
    \centering\vspace*{-25pt}
    \includegraphics[width=5.5cm]{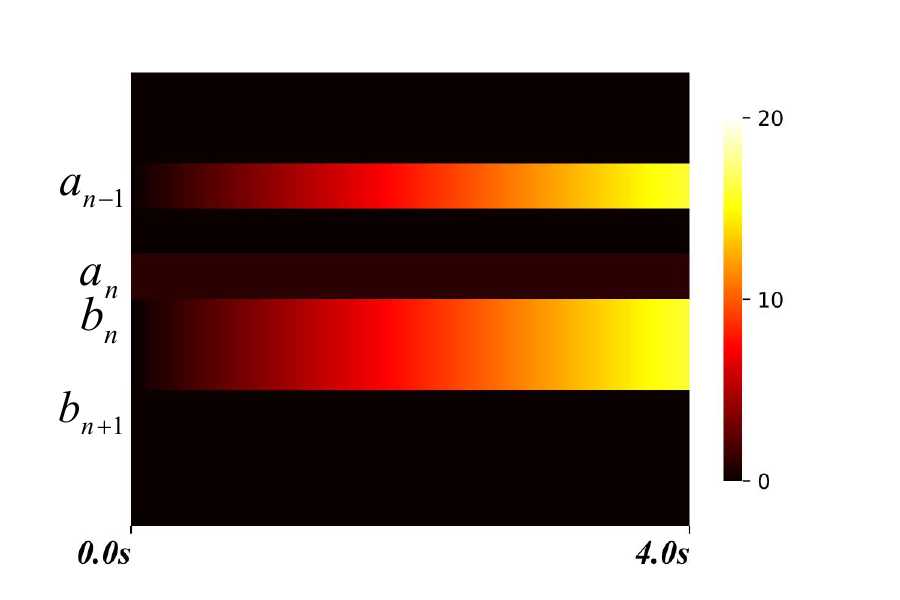}\vspace*{0pt}
    \end{minipage} \\
    \midrule[0.5pt]
    \begin{minipage}[m]{.25\textwidth} \vspace{40pt} 2ndEP2s: \\ \vspace{20pt}
     $(\tilde{H}-\lambda)^2(\tilde{H}+\lambda)^2$ 
    \end{minipage} %
   \hfill
   \begin{minipage}[m]{.12\textwidth}\vspace{35pt}
   4
   \end{minipage}  & \begin{minipage}[m]
    {.3\textwidth} \vspace{10pt} $\theta_1 \neq n_1\pi$,  \\
    \vspace{5pt} 
    $\abs{t_1} \neq \abs{t_2}$
      \vfill 
    \includegraphics[width=4cm]{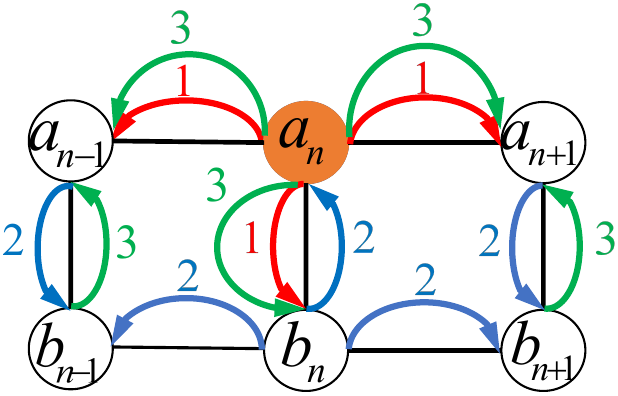}\vspace*{-32pt}
    \end{minipage} & \begin{minipage}[t]{.35\textwidth}
    \centering\vspace*{-25pt}
    \includegraphics[width=5.5cm]{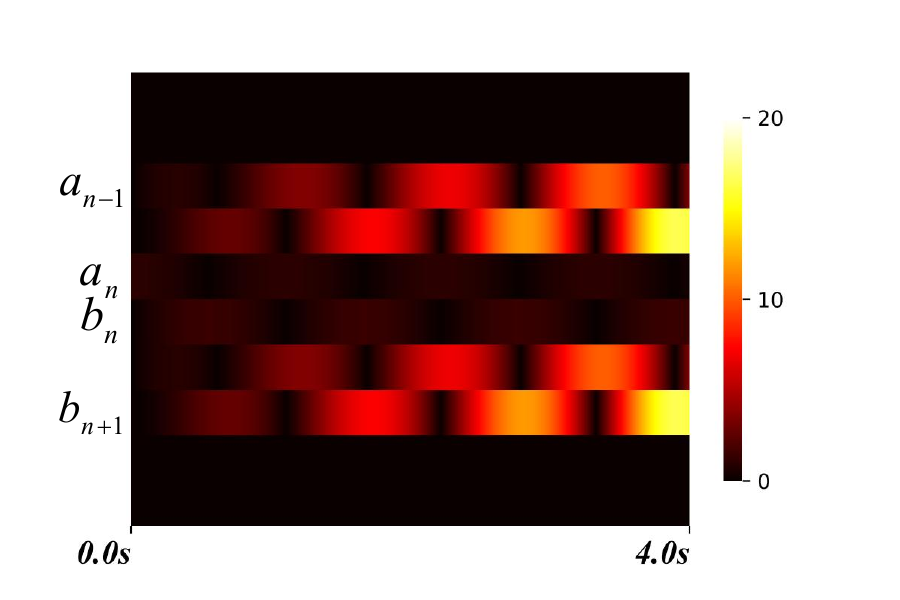}\vspace*{0pt}
    \end{minipage} \\
    \midrule[0.5pt]
   \begin{minipage}[m]{.25\textwidth} \vspace{40pt} EP4s: \\ \vspace{20pt}
    $\tilde{H}^4$
     \end{minipage} %
   \hfill
   \begin{minipage}[m]{.12\textwidth} \vspace{35pt}
   4
   \end{minipage} & \begin{minipage}[m]{.3\textwidth} \vspace{10pt}
   $\theta_1   \neq n_1 \pi$,\\
    \vspace{5pt}  $\abs{t_1} = \abs{t_2}$
    \vfill 
    \includegraphics[width=4cm]{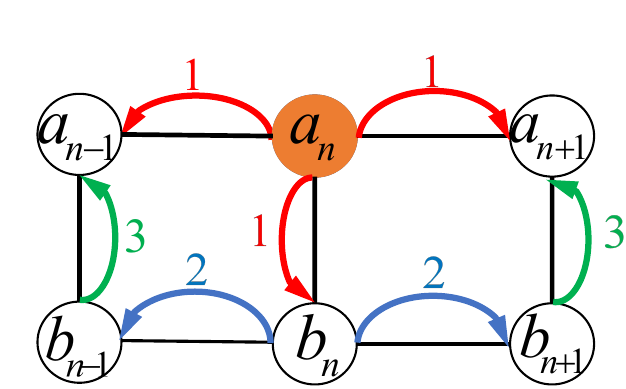}\vspace*{-32pt}
    \end{minipage} & \begin{minipage}[t]{.35\textwidth}
    \centering\vspace*{-25pt}
    \includegraphics[width=5.5cm]{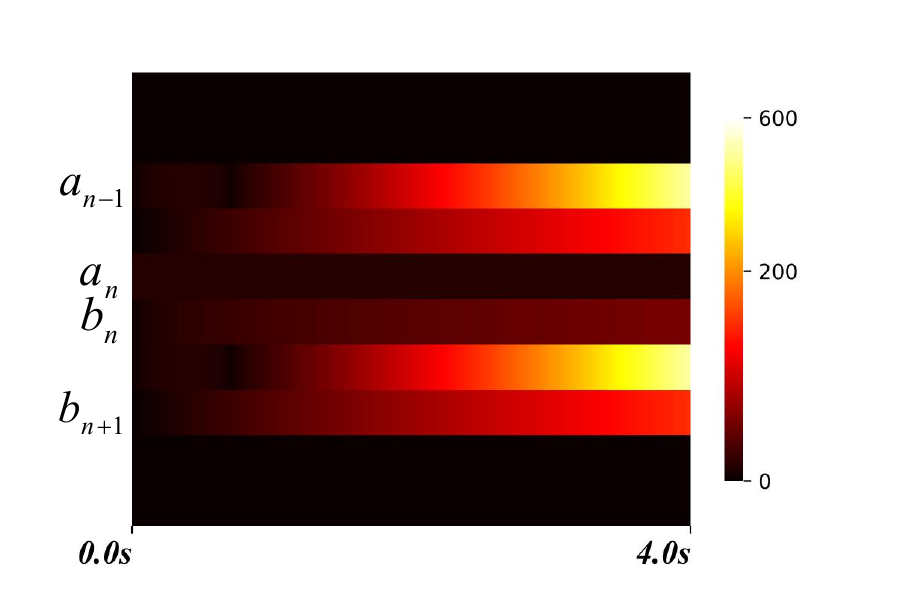}\vspace*{1pt}
    \end{minipage} \\
   \toprule[1pt]
  \end{tabular}
\end{table*}

Having established the correspondence between the minimal polynomial in mathematic and the local range of AB cage in physics, now we can use them to clarify the degeneracy type of the DFBs by comparing the specified form of the minimal polynomial and the annihilating polynomial, as shown in Table~\ref{table1}. The excitations in different kinds of AB cage can generally transition differently with different local range, as visually presented in the third column in Table~\ref{table1}. Since the leftward transition is symmetric about the rightward transition, we here only focus on the evolution of $\left[a_{n-1},b_{n-1},a_{n},b_{n}\right]$ with the initial state $\ket{\psi}_e=[0,0,\gamma_{a_n},0]$ (the third column in Table.~\ref{table1}), where the arrow with number $n$ indicates the $n$-th transition.

In the first two rows in Table~\ref{table1}, the probability of excitation transition from $a_n$ to $b_{n \pm 1}$ is zero. The transition probability $U_0U_{\downarrow}+U_{\downarrow}U_0=0$, leading to $\theta_1=n_1 \pi$, $t_1\cos\theta_1=t_2\cos\theta_2$ with $n_1 \in \mathbb{Z}$.  The corresponding state $H^n \ket{\psi}_e$ after $n$-th transition is denoted by $\ket{\psi}_n$ and 
\begin{align}
    &\ket{\psi}_1=[U_0\gamma_{a_n},0,0,U_{\downarrow}\gamma_{a_n}]; \\ \nonumber
    &\ket{\psi}_2=[0,0,U_{\downarrow}U_{\uparrow}\gamma_{a_n},0].
\end{align}
With $U_{\uparrow}U_{\downarrow}=(t_1^2-t_2^2)I_2=\lambda^2 I_2$ and $\abs{t_1} \neq \abs{t_2}$ in the first row in Table.~\ref{table1}, the second state comes back to the position of initial state, $H^2\ket{\psi}_e = \lambda^2 \ket{\psi}_e $. It will come to the same result when the excitation is at $b_n$. Thus for the single excitation at $a_n$, or $b_n$, or their linear combination, the largest local range is $2$. The minimal polynomial with order $2$ is
\begin{align}
\label{general annihilating polynomial}
    (\tilde{H}-\lambda I)(\tilde{H}+\lambda I),
\end{align}
which has two different single roots. Consider that each root is degenerate, therefore the system must have DP2s at eigenvalues $E=\pm \lambda$. The excitation transition is further verified by the numerical results of dynamic evolution in the fourth column in the Table.~\ref{table1}, where the initial excitation $ \gamma_{a_n} = [  1/\sqrt{2},1/\sqrt{2}  ]^{\intercal}$ recovers to initial state after twice transitions. These numerical results also show exponential growth intensity at each sites, cause by positive imaginary eigenvalue $E= + \lambda$, since  $\theta_1 = n_1 \pi$, $t_1 = t_2 \cos{\theta_2}$, $\abs{t_1} < \abs{t_2}$  and  $\lambda$ is a imaginary number. 

Different from the above case by only one condition $\abs{t_1}=\abs{t_2}$, the minimal polynomial $\tilde{H}^2=0$, implying that the emergence of EP2s at $0$. It is the first kind of EP2s, denoted by 1stEP2s, corresponding to the second row in the Table.~\ref{table1}. In this case, the transition is unidirectional. It stops after one transition at other sites instead of returning back the initial excitation position. The corresponding dynamical evolution shows that only the intensity at three sites $|\gamma_{a_{n\pm 1}}|^2$ and $|\gamma_{b_n}|^2$  near the excitation site grows linearly with the time due to the existence of EP2s, which aligns with the finding that the existence of EPns leads to $n$-order polynomial increase in the intensity of the $n$ generalized eigenstates~\cite{WOS:000428961400003}.

The other two rows in the Table.~\ref{table1} show two degeneracy types, where the  excitations at $a_n(b_n)$ can reach to the sites $b_{n \pm 1}(a_{n\pm 1})$ with $t_1\cos\theta_1=t_2\cos\theta_2$, $\theta_1 \neq n_1 \pi$.  When $\abs{t_1} \neq \abs{t_2}$ (the third row in Table.~\ref{table1}), with the initial excitation $\ket{\psi}_e$ at $a_n$, the excitation transition allows transition states including
\begin{align}
    &\ket{\psi}_1=[U_0\gamma_{a_n},0,0,U_{\downarrow}\gamma_{a_n}]; \\ \nonumber
    &\ket{\psi}_2=[0,(U_0U_{\downarrow}+U_{\downarrow}U_0)\gamma_{a_n},\lambda^2\gamma_{a_n},0];\\ \nonumber
    &\ket{\psi}_3=[(2\lambda^2 U_0+U_{\uparrow}U_0 U_{\downarrow})\gamma_{a_n},0,0,\lambda^2 U_{\downarrow} \gamma_{a_n}].
\end{align}
Along with the $\ket{\psi}_e$, there are four linearly independent states. The largest local range of such excitation is equal to the order of annihilating polynomial and thereby the minimal polynomial must be the annihilating polynomial. 

Note that $\abs{t_1} \neq \abs{t_2}$ and $\lambda \neq 0$, the minimal polynomial has two different double roots and thus the system also has EP2s at $\pm \lambda$. Different from the previous 1stEP2s type with largest local range of $2$, here the EP2s denoted by 2ndEP2s type possesses largest local range of $4$. The heatmap of dynamical evolution also shows the intensity at sites ${a_n, b_n}$ 
and ${a_{n \pm 1}, b_{n \pm 1}}$ oscillates with frequency $\lambda$, while the latter two sites are also accompanied by growth over time.

As for $\abs{t_1}=\abs{t_2}$, the minimal polynomial has a quadruple root and thus the system has EP4s at $0 $,  which corresponds to the fourth row in the Table.~\ref{table1}. All the transitions are unidirectional and excitation transition will end at the third transition state. The heatmap of dynamical evolution also shows that the intensity at the initial site keeps constant while intensity at $b_n$, $b_{n\pm 1}$, and $a_{n\pm 1}$ grows with time $t$ due to the existence of EP4s.

It should be noted that  our model incorporates two distinct quadratic interactions:  $a_n^{\dagger}a_{n+1}^{\dagger}+H.c.$ and $a_n^{\dagger}b_n^{\dagger}+H.c.$. In fact, the non-Hermitian AB cage can still be maintained when the second type quadratic interaction type is absent under specific parameters $t_2=0\neq t_1$ and $\theta_1=(2n_1+1)\pi/2$. However, this actually belongs to a special case in the third row of the Table~\ref{table1}, where the degeneracy type is limited to 2ndEP2s. Thus, the coexistence of two types of quadratic interaction endows the flat band with more degeneracy types.

In summary, we still use minimal polynomial to determine the degeneracy type of DFBs, while the minimal polynomial is obtained by the transfer matrix in two steps. The first step is to obtain the annihilating polynomial derived from the transfer matrix, whose factors include the minimal polynomial. The second step is to determine the local range, and hence the order of the minimal polynomial, by considering the transition of possible excitation through transfer matrix. Our results also suggest the AB cage with higher-order EPs generally  has a larger local range.

\section{The construction of DFBs with $2N$ order Exceptional Points of Degeneracy (EP2Ns)}
\begin{figure}[ht]
    \centering
    \includegraphics[width=0.48\textwidth]{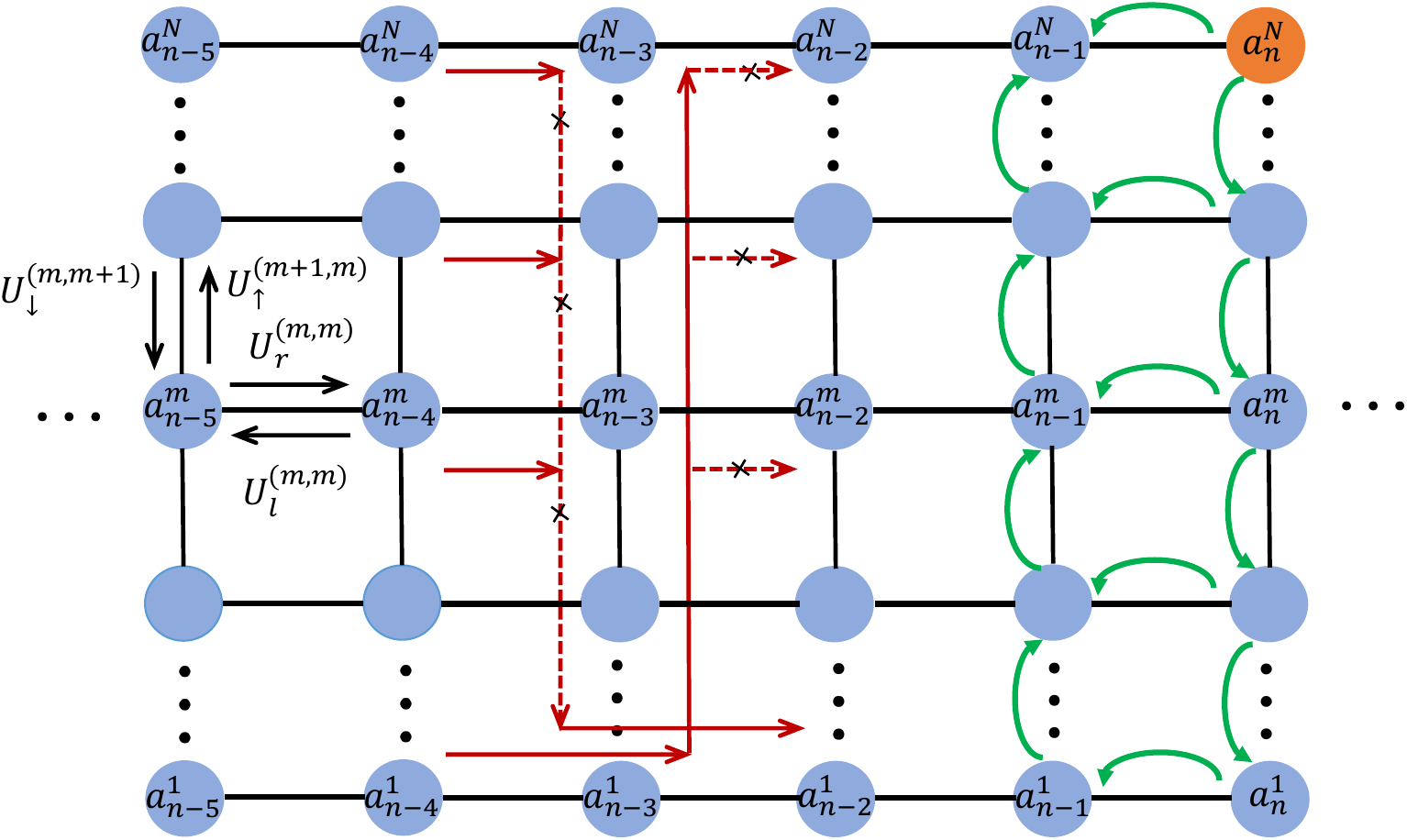}
    \caption{The realization of AB cage with $2N$-order EPs in $N$ coupled bosonic Kitaev-Majorana chains. Left part: the transfer matrices along the corresponding directions are denoted by the black arrows. Middle part: the conditions for the formation of AB cage match those for the prohibited red propagation paths, which cause the excitation at the $\left(n-4\right)$-th column to be localized between the $\left(n-5\right)$-th column and $\left(n-3\right)$-th column. Right part: the largest local range among possible excitations, illustrated by the allowed transitions (green lines).}
    \label{NEP lattice model}
\end{figure}

In this section, we utilize the local range to construct DFBs with EP2Ns in coupled $N$ bosonic Kitaev-Majorana chains, as illustrated in Fig.~\ref{NEP lattice model}. To distinguish hopping within line and between lines, the transfer matrices propagating to the right or left along the chain $m$ are denoted by $U_r^{(m,m)}$ and $U_l^{(m,m)}$, 

\begin{align}
    U_r^{(m,m)}=U_l^{(m,m)}=t^{m,m} \mqty[1&1\\-1& -1] \propto U_0.
\end{align}
Here, the conjugated coupling and number non-conservation coupling are designed to have the same strength $t^{m,m}$, to meet the conditions for forming the flat band within a single chain.  $U_{\downarrow(\uparrow)}^{(m,m+1)}$ represents the transfer matrices between the chain $m$ and $m+1$,
\begin{align}
    &U_{\uparrow}^{(m+1,m)} = \smqty[&t_1^{m,m+1}\mathrm{e}^{\mathrm{i}\theta_1^{m,m+1}} &t_2^{m,m+1}\mathrm{e}^{\mathrm{i}\theta_2^{m,m+1}} \\ &-t_2^{m,m+1}\mathrm{e}^{-\mathrm{i}\theta_2^{m,m+1}}  &-t_1^{m,m+1}\mathrm{e}^{-\mathrm{i}\theta_1^{m,m+1}}];\nonumber\\  
    &U_{\downarrow}^{(m,m+1)} = \smqty[&t_1^{m,m+1}\mathrm{e}^{-\mathrm{i}\theta_1^{m,m+1}} &t_2^{m,m+1}\mathrm{e}^{\mathrm{i}\theta_2^{m,m+1}} \\ &-t_2^{m,m+1}\mathrm{e}^{-\mathrm{i}\theta_2^{m,m+1}}  &-t_1^{m,m+1}\mathrm{e}^{\mathrm{i}\theta_1^{m,m+1}}],
\end{align}
with conjugated coupling $t_1^{m,m+1}\mathrm{e}^{\mathrm{i}\theta_1^{m,m+1}}$ and number non-conservation coupling
$t_2^{m,m+1}\mathrm{e}^{\mathrm{i}\theta_2^{m,m+1}}$. 
In the system with coupled two chains as discussed before, the additional prohibited propagation paths in Fig.~\ref{fig2} require the flat bands conditions include \(U_0 U_{\uparrow} U_0 = U_0 U_{\downarrow} U_0 = 0\). Similarly, to maintain the flat bands in the system with $N$ coupled chain, we require red propagation paths in Fig.~\ref{NEP lattice model} to be prohibited. The corresponding transfer matrices should satisfy 
\begin{align}
\label{eq-NEP path}
&U_0 U_{\downarrow}^{(m-x,m-x+1)} \cdots U_{\downarrow}^{(m-1,m)} U_0 \nonumber\\
= &U_0 U_{\uparrow}^{(m,m-1)} \cdots U_{\uparrow}^{(m-x+1,m-x)} U_0=0
\end{align}
for arbitrary $m$ and $x$ with $x<m$.

To satisfy these conditions, a straightforward choice is to ensure
\begin{align}
    U_{\downarrow}^{(m-1,m)} U_0 = 0;\quad U_0 U_{\uparrow}^{(m,m-1)}  = 0.
\end{align}
This requires $t_1^{m-1,m} = t_2^{m-1,m}$ and $\theta_1^{m-1,m} = -\theta_2^{m-1,m} \neq n_1 \pi$. These conditions precisely align with the requirement for combining two second-order EPs into a fourth-order EP in the double-chain system, resulting in arbitrary excitation at the $\left(n-4\right)$-th column to be localized between adjacent the $\left(n-5\right)$-th column and the $\left(n-3\right)$-th column. Meanwhile, $t_1^{m-1,m} = t_2^{m-1,m}$ results in $U_{\downarrow}^{(m-1,m)}U_{\uparrow}^{(m-1,m)}=0$. Along with $U_r^{(m,m)}U_l^{(m,m)}=0$, each path in the system is unidirectional.  

In order to determine the degeneracy type of this system, we need to find the annihilation polynomial of the system and determine the largest local range of the excitations in such AB cage. Now we consider an excitation $\ket{\psi}_e$ at site $a_n^N$ (the orange point in Fig.~\ref{NEP lattice model}) with transition paths shown in Fig.~\ref{NEP lattice model}. Since the leftward
transition is symmetric about the rightward transition, we here only consider the occupation of the transition states in the $n$-th and $(n-1)$-th columns. 

When the number of transitions $m$ satisfies $1 \leq m \leq N - 1$, the state after the $m$-th transition can be written as a superposition of different paths,
\begin{align}
\label{mtransition}
 &\ket{\psi}_m = \prod_{x=1}^m U_{\downarrow}^{(N-x,N-x+1)} \ket{\psi}_e \nonumber + [\cdots  U_{l}^{(y,y)} \cdots]\ket{\psi}_e.\\ 
\end{align}
The first term corresponds to the transitions that continuously occupy new site in the $n$-th column. The second term is a simplified notation for the transition to the $(n-1)$-th column, which must involves a leftward  propagation $U_{l}^{(y,y)}$ with $y \geq 1$. As the number of transitions increases, the state in the $n$-th column will shift downwards one by one. Therefore, when $1 \leq m \leq N - 1$, the states $\ket{\psi}_m$ are independent of each other, since they occupy different sites in the $n$-th column. The $N$-th transition results in multiplying $U_l^{(1,1)}$ in front of the first term. Therefore, for $N  < m \leq 2N - 1$, the state becomes 
\begin{align}
 \ket{\psi}_m &= \prod_{k=1}^{k=m-N} U_{\uparrow}^{(k+1,k)}U_l^{1}\prod_{x=1}^{N-1} U_{\downarrow}^{(N-x,N-x+1)} \ket{\psi}_e \\ \nonumber 
 & + [\cdots  U_{l}^{(y,y)}\cdots]\ket{\psi}_e ,
\end{align}
with two terms corresponding to the evolution of two terms in Eq.~\eqref{mtransition}. To determine independent states, we can still focus on the first term. It can be seen that the states in the $(n-1)$-th column shift upwards one by one, thus contributing $N-1$ independent states. These states are also linear independent from previous states, due to no occupation on the $n$-th column. Therefore, after $(2N - 1)$ times transitions, the set $\{\ket{\psi}_e, \ket{\psi}_1, \cdots, \ket{\psi}_m\}$ consists of $2N$ independent states, and the local range of the excitation $\ket{\psi}_e$ can be $2N$. It is also worth noting that those transitions transfer the state from $a_n^N$ to $a_{n-1}^N$, and the probability of further transitions is zero, i.e., $ H^{2N} \ket{\psi}_e = 0.$

For other excitations at $a_n^x$, after $N + x - 1$ transitions, the state will be located at $a_{n-1}^N$ and satisfy $\tilde{H}^{N + x } \ket{\psi}_e = 0$.
Thus, for any linear combination of excitations on these lattice sites, the equation $\tilde{H}^{2N} \ket{\psi}_e = 0$ always holds. Therefore, the polynomial
\begin{align}
  \label{eq-2Nminimalpoly}
   \tilde{H}^{2N},  
\end{align}
is the annihilation polynomial of $\tilde{H}$. Since the system has an excitation with a local range of $2N$, the degree of the minimal polynomial should be greater than or equal to $2N$. Hence, the minimal polynomial of the system is given by equation~\eqref{eq-2Nminimalpoly}, which has a degenerate root of multiplicity $2N$ at zero. Therefore, the system exhibits a flat band with an EP2Ns degeneracy.

\section{Conclusions}
In conclusion, we propose a non-Hermitian AB cage in the bosonic BdG systems. Such AB cage is manifested as DFBs, where the degeneracy type can be flexibly controlled by the system parameters. The non-Hermitian nature in the BdG systems without introducing gain or loss and intrinsic symmetry facilitate the experimental realization. For example, the parametric amplification process in superconducting quantum circuits can be effectively implemented and regulated~\cite{PhysRevLett.103.147003,wilson2011observation}, which can serve as a candidate platform for the non-Hermitian AB cage. Theoretically, we build the correspondence among the degeneracy type of DFBs, the minimal polynomial, and the transfer matrix, and according to this, we design DFBs with arbitrarily high degeneracy.  Our results provide theoretical guidance for designing the non-Hermitian AB cages and laying foundation on their dynamical properties. Currently, our work is limited to one-dimensional lattice systems, but our approach is expected to be extended to higher-dimensional systems, offering diverse platforms for studying strongly correlated physics.

This work is supported by the
Natural Science Foundation of Hunan Province ((Grant No. 2024JJ6011) and Innovation Program for Quantum
Science and Technology. (Grant No. 2021ZD0302300)

\appendix

\section{GAUGE INVARIANT NON-ABELIAN WILSON LOOP}
\label{Appendix-A}

\begin{figure}[h]
    \centering
    \includegraphics[width=0.4\textwidth]{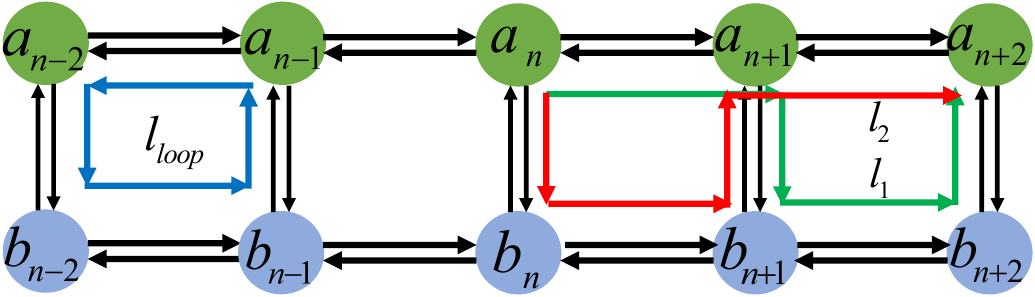}
    \caption{The illustration of gauge-invariant Wilson loop and the sufficient and necessary conditions for the forming of AB cage. Left part: the blue path $l_{loop}$ is a closed loop for excitation transition, and the trace of the link along such loop corresponds to a gauge-invariant Wilson loop. Right part: the excitations transition along the red path $l_1$ and green path $l_2$ to further sites when the transition along the chain is forbidden by the condition $J=t$. To form the AB cage, the excitation should only transition along such paths for finite periods.  }
    \label{fig.s1}
\end{figure}

Here, we explain that the chosen gauge can always ensure the phases only appear on the $t_1$ and $t_2$, which is guaranteed by the gauge-invariant Wilson loop of the system. 
The particle and hole freedoms on a single site can be expressed by a two-component operator as $ \gamma_{a_n(b_n)} = [  \alpha_n(\beta_n)  , \alpha^{\dagger}_n(\beta^{\dagger}_n)   ]^{\intercal}$. According to the equation~\eqref{eq-non-abelian} in the main text, the effective Hamiltonian can be written as
\begin{align}
    \tilde{H}_{eff} = \sum_{x_n,y_m} \gamma^{\dagger}_{x_n} U_{x_n,y_m}\gamma_{y_m},
\end{align}
with $x_n,y_m$ traversing all sites. This can be viewed as an  lattice Hamiltonian in the
presence of a U(2) gauge field $\mathbf{A}$~\cite{makeenko2009brief}. Here, the transfer matrix has the form of 
\begin{align}
    U_{x_n,y_m}=J_{x_n,y_m} P \exp \left[i \int_{[y, m]}^{[x, n]} d \mathbf{x} \cdot \mathbf{A}(\mathbf{x})\right],
\end{align}
which is the link of the sites $x_n$ and $y_m$ multiplied by the uniform  hopping strength $J_{x_n,y_m}$ between the linked sites. Here, $P$ is the path-order operator. For a given path that forms a loop, the gauge-invariant Wilson loop is defined as the trace of the product of link along the loop
\begin{align}
    W_{loop} = Tr(P \exp \left[i \oint d \mathbf{x} \cdot \mathbf{A}(\mathbf{x})\right]).
\end{align}

Let's consider an excitation starting from site $a_{n-2}$ and returning to its original position along the blue path $l_{loop}$ in Fig.~\ref{fig.s1}. The Wilson loop along the path is
\begin{align}
    W_{loop} = & Tr(U_{loop}/J_{loop)}.
\end{align}
Here, $J_{loop}= J_{a_{n-2},a_{n-1}} J_{a_{n-1},b_{n-1}} J_{b_{n-1},b_{n-2}} J_{b_{n-2},a_{n-2}}$ is the hopping strength along the loop, and $U_{loop}=U_l U_{\uparrow} U_r U_{\downarrow}$ is the corresponding transfer matrix for this loop  with the matrix form
\begin{align}
    U_{loop} = \mqty[u_{1,1}& u_{1,2} \\ u_{2,1}& u_{2,2} ].
\end{align}
The matrix elements of $U_{loop}$ are calculated as follows,
\begin{widetext}
\begin{align}
    &u_{1,1} = [t_1^2\qty(1+\mathrm{e}^{-\mathrm{i}\Phi_1})+t_1t_2\qty(\mathrm{e}^{\mathrm{i}\Gamma_1}+\mathrm{e}^{-\mathrm{i}\Gamma_2}+\mathrm{e}^{\mathrm{i}\Gamma_1}+\mathrm{e}^{\mathrm{i}\Gamma_2})+t_2^2\qty(1+\mathrm{e}^{-\mathrm{i}\Phi_2})]tJ , \notag \\ 
    &u_{1,2} = [t_1t_2\qty(2+\mathrm{e}^{-\mathrm{i}\Phi_1}+\mathrm{e}^{-\mathrm{i}\Phi_2})+t_2^2\qty(\mathrm{e}^{\mathrm{i}\Gamma_1}+\mathrm{e}^{\mathrm{i}\Gamma_2})+t_1^2\qty(\mathrm{e}^{\mathrm{i}\Gamma_1}+\mathrm{e}^{-\mathrm{i}\Gamma_2})]tJ \mathrm{e}^{\mathrm{i}(\theta_1+\theta_2)},  \notag \\
    &u_{2,1} = [t_1^2\qty(1+\mathrm{e}^{-\mathrm{i}\Phi_1})+t_1t_2\qty(\mathrm{e}^{\mathrm{i}\Gamma_1}+\mathrm{e}^{-\mathrm{i}\Gamma_2}+\mathrm{e}^{\mathrm{i}\Gamma_1}+\mathrm{e}^{\mathrm{i}\Gamma_2})+t_2^2\qty(1+\mathrm{e}^{-\mathrm{i}\Phi_2})]tJ\mathrm{e}^{-\mathrm{i}(\eta_a+\pi)} , \notag \\ 
    &u_{2,2} = [t_1t_2\qty(2+\mathrm{e}^{-\mathrm{i}\Phi_1}+\mathrm{e}^{-\mathrm{i}\Phi_2})+t_2^2\qty(\mathrm{e}^{\mathrm{i}\Gamma_1}+\mathrm{e}^{\mathrm{i}\Gamma_2})+t_1^2\qty(\mathrm{e}^{\mathrm{i}\Gamma_1}+\mathrm{e}^{-\mathrm{i}\Gamma_2})]tJ \mathrm{e}^{-\mathrm{i}\Gamma_1}, 
\end{align}
\end{widetext}
where
\begin{align}
    &\Gamma_1=-\theta_1-\theta_2+\eta_a+\pi \nonumber \\ 
    &\Gamma_2=-\theta_1+\theta_2-\eta_b+\pi \nonumber \\ 
    &\Phi_1=2\theta_1+\eta_b-\eta_a \nonumber \\ 
    &\Phi_2=2\theta_2-\eta_a-\eta_b    
\end{align}
The Wilson loop $W_{loop}$ is invariant under a gauge transformation~\cite{makeenko2009brief}, which requires \( \Phi_1 \), \( \Phi_2 \), \( \Gamma_1 \), and \( \Gamma_2 \) are gauge-invariant. Note that \( \Gamma_1 + \Gamma_2 = -\Phi_1 \) and \( \Gamma_1 - \Gamma_2 = -\Phi_2 \) and therefore \( \theta_1 \), \( \theta_2 \), \( \eta_a \), and \( \eta_b \) only need to satisfy the third and fourth constraints in the above equations. When we take the gauge transformation  \(\alpha_n \rightarrow \alpha_n \mathrm{e}^{\mathrm{i}\varphi_a}\) and \(\beta_n \rightarrow \beta_n \mathrm{e}^{\mathrm{i}\varphi_b}\),
\begin{align}
    &\theta_1 \rightarrow \theta_1+\varphi_b-\varphi_a , \nonumber\\
    &\theta_2 \rightarrow \theta_2-\varphi_b-\varphi_a , \nonumber\\
    &\eta_a \rightarrow \eta_a-2\varphi_a, \nonumber \\
    &\eta_b \rightarrow  \eta_b-2\varphi_b. 
\end{align}
It can be clearly seen that $\Phi_1$ and $\Phi_2$ don't change under the gauge transformation. Therefore, we can always perform a gauge transformation to apply all phases to the terms representing the coupling strengths between the two chains, such that \( \eta_a = \eta_b = 0 \), \( \theta_1 = \Phi_1/2 \), and \( \theta_2 = \Phi_2/2 \). Our subsequent calculations are based on this chosen gauge.

\section{FLAT BAND CONDITIONS}
\label{Appendix-B}
The flat band conditions can be simplified as two limited propagation paths in the main text, as demonstrated below. The localization of excitation in single chain $``a"$ or $``b"$ requires $(U_r)^n = (U_l)^n =0 $ with integer $n$. For the case of $n=1$, this corresponds to $U_r = U_l =0$, implying that the sites are uncoupled, and therefore can be ignored. The dimension of $U_r = U_l$  constrains $n$, such that the case  $(U_r)^2 = (U_l)^2 \neq 0 $ but $(U_r)^3 = (U_l)^3 =0 $ cannot occur. Therefore, $n=2$, which leads to $J=t$ and $U_r=U_l=U_0$. While for coupled two chains under conditions $U_r=U_l=U_0$ and $U_r^2=U_l^2=0$, there are additional possible transition paths alternatively propagating two legs, such as $a_n \rightarrow b_{n+1} \rightarrow a_{n+2} \rightarrow b_{n+3} \rightarrow a_{n+4}$ periodically, as shown in Fig.~\ref{fig.s1}. With $U_{\uparrow}U_{\downarrow}=(t_1^2-t_2^2)I_2$, the transfer matrix in one period from $a_n$ to $a_{n+2}$ is
\begin{align}
    U_{a_{n+2},a_{n}} = &(U_{\uparrow}U_0+U_0 U_{\uparrow})  (U_{\downarrow}U_0+U_0 U_{\downarrow}) \notag \\
    =&U_{\uparrow}U_0U_{\downarrow}U_0+U_0U_{\uparrow}U_0 U_{\downarrow},
\end{align} which is equal to the superposition of path $l_1$ (the green path in Fig.~\ref{fig.s1}) with transfer matrix $U_1=U_{\uparrow}U_0 U_{\downarrow}U_0$ and path $l_2$ (the red path in Fig.~\ref{fig.s1}) with  $U_2=U_0 U_{\uparrow} U_0 U_{\downarrow}$. 

For the formation of AB cage, the transition along the paths $l_1$, $l_2$ can only sustain finite periods with $(U_{a_{n+2},a_{n}})^m=0$. Since the dimension of the $U_{a_{n+2},a_{n}}$ is $2$, here $m$ also can not be larger than 2. $U_{a_{n+2},a_{n}}$ has the form of
\begin{small}
\begin{align}
   U_{a_{n+2},a_{n}}=Jt\mqty[(t_1\cos\theta_1-t_2\cos\theta_2)^2& 0\\0 & (t_1\cos\theta_1-t_2\cos\theta_2)^2]. 
\end{align}
\end{small}
It can be seen that $(U_{a_{n+2},a_n})^2=0$ is equivalent to the equation~\eqref{eq-transion-path-between-chains} in the main text. Thus the conditions $t_1\cos\theta_1=t_2\cos\theta_2$ and $J=t$  are both the necessary and sufficient conditions for the existence of an AB cage in our system.

\section{The Local Range of AB Cage and the Minimal Polynomial }
\label{Appendix-C}

There always exists an excitation, whose minimal polynomial (Eq.~\eqref{eq-repeatpath}) is identical to the minimal polynomial of $\tilde{H}$ [Eq.~\eqref{eq-minipoly}] and the local range of the the AB cage is equal to $l$~\cite{curtis2012linear}.  To prove this, we firstly utilize a mathematical corollary, which states that a matrix with minimal polynomial~\eqref{eq-minipoly} can be similar to a block diagonal matrix containing the following matrix as diagonal block~\cite{hungerford2012algebra,curtis2012linear,jacobson2012basic},
\begin{align}
\label{eq-diagonal block}
   C = \left(\begin{array}{ccccc}
0 & 0 & \cdots & 0 & -c_0 \\
1 & 0 & \cdots & 0 & -c_1 \\
0 & 1 & \cdots & 0 & -c_2 \\
\vdots & \vdots & & \vdots & \vdots \\
0 & 0 & \cdots & 0 & -c_{l-2} \\
0 & 0 & \cdots & 1 & -c_{l-1}
\end{array}\right).
\end{align}
Applying the above matrix to a specified state $\ket{e_k}$ with $l$ components, in which the subscript $k$ indicates that only the $k$-th component is 1 with all other components being zero, we have
\begin{equation}
\begin{cases}
\ket{e_{k+1}}=C \ket{e_k}, \quad \text{for} \quad  k<l; \\
C \ket{e_l} = -c_0 \ket{e_1} -c_1 \ket{e_2} - \cdots -c_{l-1}\ket{e_l}, \quad \text{for} \quad  l.
\end{cases} 
\end{equation}
The second equation can be simplified as 
\begin{align}
    C^l \ket{e_1}+\sum_{n=0}^{l-1}c_n C^n \ket{e_1}=0.
\end{align}
The collection of states $\{\ket{e_1}, \cdots \ket{e_l}\}$ forms a cyclic space with dimension $l$, which is the basis of the diagonal block~\eqref{eq-diagonal block}, and the minimal polynomial of $\ket{e_1}$ is the same as the minimal polynomial~\eqref{eq-minipoly} of $\tilde{H}$. Therefore, the local range of excitation  $\ket{e_1}$ can be $l$ and the local range of the AB cage can take its upper bound, i.e., the order of minimal polynomial.

\section{ANNIHILATING POLYNOMIAL OF $\tilde{H}$}
\label{Appendix-D}
To obtain the annihilating polynomial, we first present the form of $\tilde{H}^n$ with different orders. Without loss of generality, we only consider the case where the excitation is located at $a_n$, in the following discussion. The excitation at site $ a_n $ can propagate to $ b_n$, $ a_{n+1} $, and $ a_{n-1}$ as 
\begin{align}
    \tilde{H}_{b_n,a_n}=U_{\downarrow}, \quad \tilde{H}_{a_{n+1},a_n}=\tilde{H}_{a_{n-1},a_n}=U_0
\end{align}
during the first transition. 
Note that all other sub-matrices of $\tilde{H}$ not mentioned are zero matrices. During the second transition, the excitation can reach to sites $b_{n+1}$, $b_{n-1}$ or come back to $a_n$. This results in the following nonzero sub-matrices
\begin{align} {\tilde{H}^2}_{a_n,a_n}&=U_{\uparrow}\tilde{H}_{b_n,a_n}=(t_1^2-t_2^2)I_{2};\\
    {\tilde{H}^2}_{b_{n+1},a_n}&={\tilde{H}^2}_{b_{n-1},a_n} =U_{\downarrow}\tilde{H}_{a_{n-1},a_n}+U_0\tilde{H}_{b_n,a_n} \notag \\    &=U_{\downarrow}U_0+U_0U_{\downarrow}.
\end{align}

After three transitions, the excitation returns to sites  $b_n$,  $a_{n+1}$, and  $a_{n-1}$, yielding
\begin{align}
    {\tilde{H}^3}_{b_n,a_n}&=U_{\downarrow}{\tilde{H}^2}_{a_n,a_n}+U_0{\tilde{H}^2}_{b_{n-1},a_n}+U_0{\tilde{H}^2}_{b_{n+1},a_n} \notag \\
    &=(t_1^2-t_2^2)U_{\downarrow};\\
    {\tilde{H}^3}_{a_{n+1},a_n}&={\tilde{H}^3}_{a_{n-1},a_n}=U_0{\tilde{H}^2}_{a_n,a_n}+U_{\uparrow}{\tilde{H}^2}_{b_{n-1},a_n} \notag \\
    &=2(t_1^2-t_2^2)U_0+U_{\uparrow}U_0U_{\downarrow}.
\end{align}

During the fourth transition, the excitation again reaches to sites $b_{n+1}$, $b_{n-1}$, and $a_n$, yielding
\begin{align}
    {\tilde{H}^4}_{a_n,a_n}&=U_{\uparrow}{\tilde{H}^3}_{b_n,a_n}+U_0 {\tilde{H}^3}_{a_{n+1},a_n}+U_0 {\tilde{H}^3}_{a_{n-1},a_n} \notag \\
    &=(t_1^2-t_2^2)^2I_{2}; \\
    {\tilde{H}^4}_{b_{n+1},a_n}&={\tilde{H}^4}_{b_{n-1},a_n}=U_{\downarrow}{\tilde{H}^3}_{a_{n-1},a_n}+U_0{\tilde{H}^3}_{b_n,a_n} \notag \\
    &=2(t_1^2-t_2^2) (U_{\downarrow}U_0+U_0U_{\downarrow}).    
\end{align}
A combination of above $\tilde{H}^n $ with different orders gives rise to the annihilating polynomial of $\tilde{H}$ as
\begin{align}
    (\tilde{H}-\lambda I)^2(\tilde{H}+\lambda I)^2=0,
\end{align}
which is the same as the result [Eq.~\eqref{periodic dynamical eigenvalues}] previously obtained using $\gamma$ matrix in momentum space.

\section{The Realization of Non-Hermitian AB cage In Superconduct Circuit}

\begin{figure}[h]
    \centering
\includegraphics[width=0.46\textwidth]{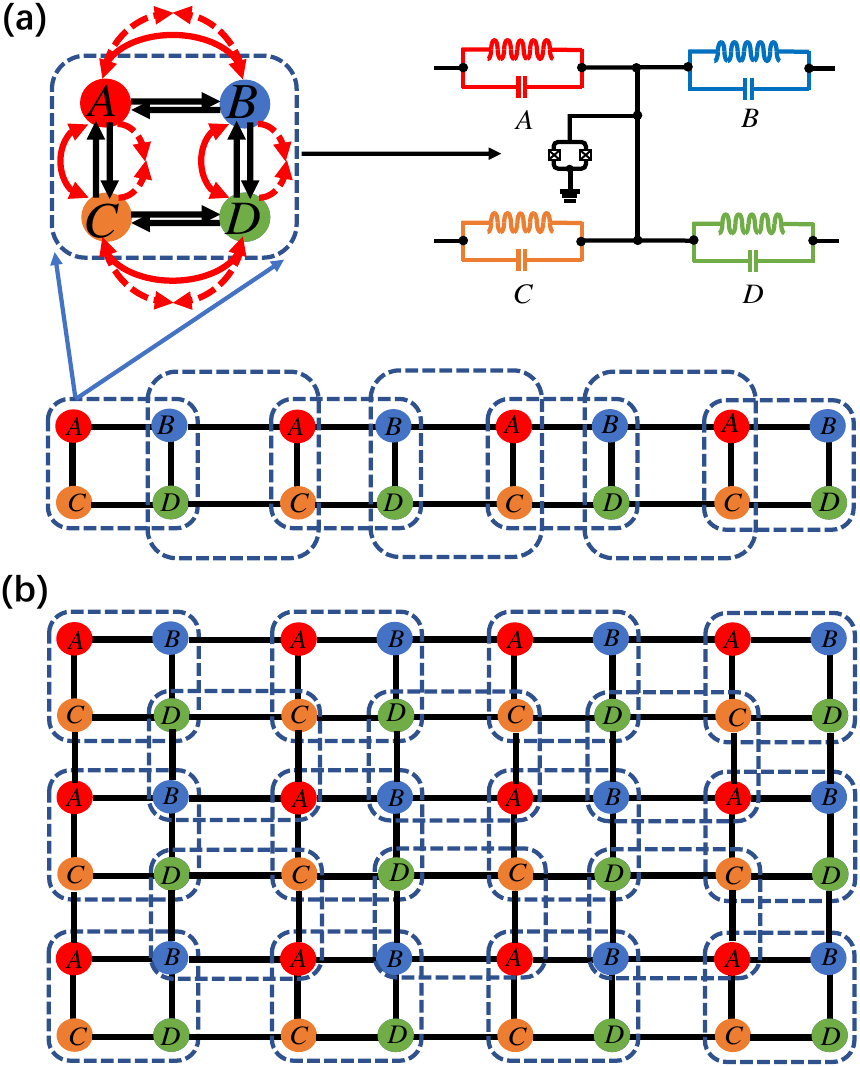}
    \caption{(a) The realization of non-Hermitian AB cage in superconducting quantum circuits with configurations similar to (a) Fig.~1 and (b) Fig.~4. The dashed box in (a) highlights the coupling between sites, where the conjugated coupling and two-boson creation/annihilation processes are denoted by the black lines, red solid/dashed lines, respectively. Here, the sites correspond to TLRs and the coupling between TLRs can be modulated via a grounding SQUID.}
    \label{circuit_lattice}
\end{figure}

Here, we propose a theoretical scheme for implementing the non-Hermitian AB cage with superconducting quantum circuits. Corresponding to coupled bosonic Kitaev-Majorana chains (Fig.~1 and Fig.~4 in the main text), similar designs can be adopted for the superconducting quantum circuits to realize non-Hermitian AB cage as shown in  Fig.~\ref{circuit_lattice}, where the sites correspond to  superconducting transmissionline resonators (TLRs)~\cite{devoret2013superconducting,you2011atomic}. The conjugated coupling and two photon creation/annihilation coupling between TLRs are exemplified by a dashed box in Fig.~\ref{circuit_lattice}(a), which contains four TLRs $A$, $B$, $C$ and $D$ with distinct fundamental frequencies ($\omega_a$, $\omega_b$, $\omega_c$, $\omega_d$). The TLRs share a common grounding node through a flux-tunable superconduct quantum interference device (SQUID)~\cite{wilson2011observation,zakka2011quantum,PhysRevLett.113.093602} characterized by Josephson energy $E_J = E_{J0}\cos(\pi\Phi_{\text{ext}}/\Phi_0)$ where $\Phi_{\text{ext}}$ denotes the externally applied flux, while $E_{J0}$ and $\Phi_0$ are the Josephson energy and flux quantum. When $E_{J0} \gg 1$, the SQUID's low impedance effectively grounds all TLRs, imposing $n\lambda/2$ standing wave modes with wavelength $\lambda$~\cite{wang2016detecting,PhysRevA.93.062319,PhysRevA.102.023524}.  In this way, such four TLRs are approximately independent with each other, and the Hamiltonian for these four TLRs can be written as
\begin{align}
    H_0 = \sum_{\eta} \hbar \omega_{\eta} \eta ^{\dagger}\eta .
\end{align}
Here, $\omega_{\eta}$ is the corresponding mode frequency of TLRs. Meanwhile, since the currents of the four TLRs converge to the ground through the SQUID, the SQUID with a small inductance  can slightly couple the four modes of TLRS with interaction Hamiltonian~\cite{PhysRevLett.113.093602,wang2016detecting},
\begin{align}
     H_{int} = A_{\omega}\cos(\omega t+\phi_{\omega})\sum_{\langle\eta,\epsilon\rangle} T_{\eta\epsilon}(\eta +\eta^{\dagger})(\epsilon +\epsilon^{\dagger}),
\end{align}
with $\eta,\epsilon\in \{a,b,c,d\}$ . $T_{\eta \epsilon}$ is the strengthen of coupling which is determined by the parameters of the TLRs and SQUID. $A_{\omega}\cos(\omega t+\phi_\omega)$ is  the AC external flux with amplitude $A_{\omega}$, frequency $\omega$ and phase $\phi_{\omega}$. Upon transformation to the interaction picture, we obtain
\begin{align}
    &H_{I}=e^{iH_0t}(H_0+H_{int})e^{-iH_0t}
    =A_{\omega}\cos(\omega t+\phi_{\omega}) \cdot \notag \\ &\sum_{\langle\eta,\epsilon\rangle} T_{\eta\epsilon}(\eta e^{-i\omega_{\eta}t}+\eta^{\dagger}e^{i\omega_{\eta}t})(\epsilon e^{-i\omega_{\epsilon}t}+\epsilon^{\dagger}e^{i\omega_{\epsilon}t}).
\end{align}

The system exhibits two distinct parametric resonance regimes~\cite{PhysRevApplied.8.024018,PhysRevLett.113.093602}. When the AC external flux frequency is equal to $\omega_{\eta}-\omega_{\epsilon}$, the parametric resonance is in frequency conversion regime which can lead to the conjugated coupling between adjacent sites, while the two photon creation/annihilation coupling can be neglected due to the rotating wave approximation.  When the AC external flux frequency is equal to $\omega_{\eta}+\omega_{\epsilon}$, the parametric resonance is in amplification regime which can lead to the two photon creation/annihilation coupling between adjacent sites, while the conjugated coupling can be neglected. 
To realize the non-Hermitian AB cage in such superconducting quantum circuits in Fig.~\ref{circuit_lattice}(a), it only requires reproducing the conditions for the flat bands in the main text by modulating the parameters. For example, one condition $J=t$ can be realized by adjusting the strength and initial phase of the AC external flux as $T_{ab}A_{\omega_a-\omega_b}=T_{ab}A_{\omega_a+\omega_b}=T_{cd}A_{\omega_c-\omega_d}=T_{cd}A_{\omega_c+\omega_d}$,  $\phi_{\omega_a-\omega_b}=\phi_{\omega_a+\omega_b}=\phi_{\omega_c-\omega_d}=\phi_{\omega_c+\omega_d}=0$.   Similarly, another condition $t_1\cos{\theta_1}=t_2\cos{\theta_2}$ is equivalent to $T_{ac}A_{\omega_a-\omega_c}\cos(\phi_{\omega_a-\omega_c})=T_{ac}A_{\omega_a+\omega_c}\cos(\phi_{\omega_a+\omega_c})$ and $T_{bd}A_{\omega_b-\omega_d}\cos(\phi_{\omega_b-\omega_d})=T_{bd}A_{\omega_b+\omega_d}\cos(\phi_{\omega_b+\omega_d})$.

\bibliography{manuscript.bib}

\end{document}